# Theoretical limits in detachment strength for axisymmetric bi-material adhesives

Farid Hoseynian Benvidi [a], Mattia Bacca [a*]

[a] *Mechanical Engineering Department, University of British Columbia, Vancouver BC V6T1Z4, Canada*

**Abstract**

Dry adhesives rely on short-ranged intermolecular bonds, hence requiring a low elastic modulus to conform to the surface roughness of the adhered material. Under external loads, however, soft adhesives accumulate strain energy, which release drives the propagation of interfacial flaws prompting detachment. The ideal adhesive is then soft but rigid. The solution to this controversial requirement is a bi-material adhesive having a soft tip, for surface conformation, and a rigid backing, for reduced strain energy release, hence, better adhesive strength. This design strategy is widely observed in nature across multiple species. However, the detachment mechanisms of these adhesives are poorly understood and quantitative analysis of their adhesive strength is still missing. Based on linear elastic fracture mechanics, we analyze the strength of axisymmetric bi-material adhesives. We observed two main detachment mechanisms, namely (i) center crack propagation and (ii) edge crack propagation. If the soft tip is sufficiently thin, mechanism (i) dominates and provides stable crack propagation, thereby toughening the interface. We ultimately provide the maximum theoretical strength of these adhesives obtaining closed form estimates for the detachment stress independent of the crack size, rendering the interface flaw tolerant.

[*] mbacca@mech.ubc.ca

**Introduction**

Adhesion has been a topic of scientific interest for decades. Nature provided numerous examples of structures that adhere through short-ranged intermolecular bonds [1–3]. Engineering applications range from fibrillar dry adhesives [4–6] inspired by spiders, beetles, and geckos [7], pressure-sensitive adhesive layers [8–10], flip-chip technology [11], skin adhesives [12,13] and hydrogel films [14]. Recent investigations revealed that the strength of these adhesives, while depending on physical properties, is also broadly correlated with geometry [15–19]. Interfacial flaws are created by manufacturing defects, surface roughness, attachment imperfection, particles and dust [20–22]. Crack propagation continues so long that the strain energy released compensates the energetic cost of interfacial debonding. It depends on the strength of the van der Waals bonds developed at the interface and can be augmented by the adoption of fibrillar geometries. Here, the strain energy stored by each fibril cannot be used to drive further detachment propagation. The maximum adhesive strength attainable from a fibrillar adhesive, however, is that of its single adhesive unit (in this case, the fibril). Hence one needs to consider the geometry of the single unit to maximize its strength. Smaller units have shown to give higher adhesive strength [7] and ultimately reach flaw insensitivity for nano-scale adhesive contacts, as observed in geckos [15].

Due to manufacturing challenges, however, the smallest prototypes of these adhesive units are commonly not smaller than tens of microns. Thus, more effort is needed to define the optimal geometry of a single adhesive unit to scale robust adhesion.

Novel designs primarily inspired by nature [23,24] to enhance adhesive strength of a single adhesive unit include mushroom-shaped fibrils [25,26], funnel-shaped fibrils [27,28] and composite pillars with a soft tip and stiff stalk [29,30]. The latter solution has proven to provide the optimal compromise between conformability to the roughness of the adhered surface, thanks to the soft tip, and adhesive strength. The adhesive strength is increased thanks to the reduction in the strain energy release rate given by the rigidity of the stalk. The bending rigidity of the pillar also helps to prevent fibril condensation, hence promoting better contact splitting at the fibrillar array level.

To formulate quantitative design principles to optimize adhesive strength, one has to identify all detachment mechanisms of the adhesive in relation to specific design parameters. Stress concentration typically occurs at the perimeter of the contact due to the geometrical singularity of the adhesive contact [31]. In this case, detachment starts from the edge of the interface, and most theoretical models developed thus far only considered this mechanism. However, experiments suggested the possibility of developing alternative detachments mechanisms such as the propagation of a central penny-shaped crack and cavitation [32,33].

In this paper we analyze the detachment strength of a axisymmetric bi-material adhesive composed of a soft adhesive tip perfectly bonded to a rigid backing and adhered to a rigid substrate. This constitutes a model system representing composite fibrils [29,30], when the stalk is rigid compared to the tip, but can also represent a broader set of applications. Our analysis evidences the predominance of two main detachment mechanisms, which control the overall strength of the adhesive. These are (i) *center crack propagation* and (ii) *edge crack propagation*. The pull-off strength of the adhesive, associated with (ii), is inversely proportional to the thickness of the soft adhesive layer [30]. This was also observed for the case of a rigid pillar detaching from an elastic adhesive layer [34]. However, as the tip becomes thinner, compared to the radius of the contact, the interface develops a stress concentration at the center of the interface. This stress concentration is inversely proportional to the tip thickness, evidencing a trade-off between the acquired strength from (ii) and the emergence of (i). We observed, however, that when center crack propagation dominates, based on the crack size and on the tip thickness, the detachment mechanism is stable. Center crack stability finally provides the theoretical limit in the detachment strength of the adhesive, making it flaw tolerant.

We base our analysis on linear elastic fracture mechanics (LEFM), and compute the mixed mode stress intensity factor and the energy release rate for both (i) and (ii) using finite element analysis.

**Mathematical model**

*Bi-material adhesive interface*

The detachment of an adhesive interface occurs when the (strain) energy release rate $G$ exceeds the work of adhesion $W_A$. Under this condition, the detachment front propagates in the form of a crack. The energy release rate is defined as,

$$G = -\frac{\partial \Pi}{\partial A} = \frac{\partial}{\partial A}(W_{ex} - \Omega) \tag{1}$$

with $W_{ex}$ the work done by an external force, and $\Omega$ the elastic strain energy and $A$ the extended crack area.

We investigate here the detachment of two layers of with the substrate being significantly stiffer than the adhesive[1]. Under this hypothesis, we can consider the substrate to be rigid, hence unable to store elastic strain energy. Based on the theory of linear elastic interfacial fracture mechanics, the asymptotic solution of the stress and strain fields near the crack tip is described as a function of the stress intensity factors and interface material properties Dundurs' elastic constants are defined as,

$$\alpha = \frac{E'_2 - E'_1}{E'_1 + E'_2}, \quad \beta = \frac{\mu_1(\kappa_2 - 1) - \mu_2(\kappa_1 - 1)}{\mu_1(\kappa_2 + 1) + \mu_2(\kappa_1 + 1)} \tag{2a}$$

where for plane strain and axisymmetric conditions,

$$E'_i = \frac{E_i}{1 - \nu_i^2}, \quad \kappa_i = 3 - 4\nu_i, \quad \text{with } i = 1,2 \tag{2b}$$

$E_i$ is the Young modulus of the two materials, $\nu_i$ is their Poisson's ratio, and $\mu_i$ is their shear modulus. The mismatch oscillation parameter is defined as,

$$\epsilon = \frac{1}{2\pi} \ln\left(\frac{1-\beta}{1+\beta}\right) \tag{3}$$

By assuming material 2 to be rigid it gets simplified to

$$\epsilon = \frac{1}{2\pi} \ln(3 - 4\nu_1) \tag{4}$$

For a nearly incompressible material ($\nu_1 \simeq 0.5$), the constants $\beta$ and $\epsilon$ approach zero. The stress intensity factor for such problems is defined as complex number

$$\boldsymbol{K} = K_I + iK_{II} \tag{5}$$

where the real part corresponds to mode I and the imaginary part to the mode II fracture [35]. both modes are simultaneously present even for uniaxial normal loading and the phase angle for mode mixing as,

$$\psi = \tan^{-1}\left(\frac{Im[\boldsymbol{K}c^{i\epsilon}]}{Re[\boldsymbol{K}c^{i\epsilon}]}\right) \tag{6}$$

which is a function of the crack length, $c$. For the case of $\epsilon = 0$, the phase angle in Eq. (6) becomes independent of $c$, giving $\psi = \tan^{-1}(K_{II}/K_I)$. The energy release rate is related to the complex stress intensity factor by

---

[1] For example, using PDMS pillars ($E = 2.6\ MPa$) on a glass substrate ($E \cong 80\ GPa$) [19]

$$G = \frac{|K|^2}{2E^* \cosh(\pi\epsilon)} \tag{7a}$$

where $|K| = \sqrt{K_I^2 + K_{II}^2}$ is the modulus of the complex stress intensity factor, and

$$\frac{1}{E^*} = \frac{1}{E_1'} + \frac{1}{E_2'} \tag{7b}$$

Considering $E_2' \to \infty$, Eq. (7b) resolves to $E^* = E_1'$. From now on, we will consider $E^* = E/(1-\nu^2)$ the plane strain modulus of the adhesive, with $E$ and $\nu$ its Young modulus and Poisson's ratio, respectively. The stress intensity factor is related to the remote normal stress $\sigma_\infty$ by

$$\boldsymbol{K} = \sigma_\infty \sqrt{c}\, \widehat{\boldsymbol{K}} \tag{8}$$

where $\widehat{\boldsymbol{K}} = \widehat{K}_I + i\widehat{K}_{II}$ is the dimensionless complex stress intensity factor.

*Finite Element Model*

The model system adopted in this investigation is sketched in Figure 1. An elastic disk of thickness $h$ and radius $R$ is perfectly bonded to a rigid backing (RB) (on top of the disk) and adhered to a rigid substrate (RS) (below the disk). Both the RB and the RS interfaces are subject to no-slip condition. While debonding does not occur at the disk-RB interface, debonding at the disk-RS interface occurs due to the presence of two cracks, $c_1$, at the center, and $c_2$, at the edge. The geometry of both cracks preserves axisymmetry, hence $c_1$ is a penny-shaped crack, and $c_2$ is a ring-shaped crack. We consider that if both $c_1$ and $c_2$ are sufficiently small compared to R, the propagation of $c_1$ is independent of $c_2$ and vice-versa. A constant remote stress $\sigma_\infty$ is applied to the RB atop the disk, which imposes a uniform displacement prescribed at the top of the elastic disk.

The geometry of the model is described by the dimensionless disk thickness, $\hat{h} = h/R$, and the dimensionless crack size $\hat{c} = c/R$, for both center crack and edge crack. The confinement of the elastic disk-shaped adhesive tip increases as $\hat{h}$ reduces.

We performed finite element (FE) analysis for a model using Abaqus with Quadratic Hybrid Axisymmetric elements, CAX8H and CAX6H to capture the incompressible behavior for a Poisson's ratios beyond 0.485. The element around the crack tip is modeled with quarter-point collapsed sided elements [36], which have a built-in stress mapping with $1/\sqrt{r}$ singularity. This allows for an accurate calculation of the energy release rate. We calculated the contour J-integral based on [37,38] with Abaqus. To simulate the rigidity of the RS, we modeled the substrate as a linearly elastic material with a modulus that is $10^9$-times that of the disk.

The FE mesh must be refined around the crack tip for better precision. Also, to limit element distortion, one needs to preserve a near-unity aspect ratio in each element. To satisfy both requirements, we adopted the spider-mesh shown in Figure 1, for locations further from the crack a coarser structured mesh was uses. To ensure the mesh independency of our results and the path independence of the contour integrals, we tested various FE models. These models included various element size and integration paths. The deviation among results with further mesh refinement ultimately was within 5%.

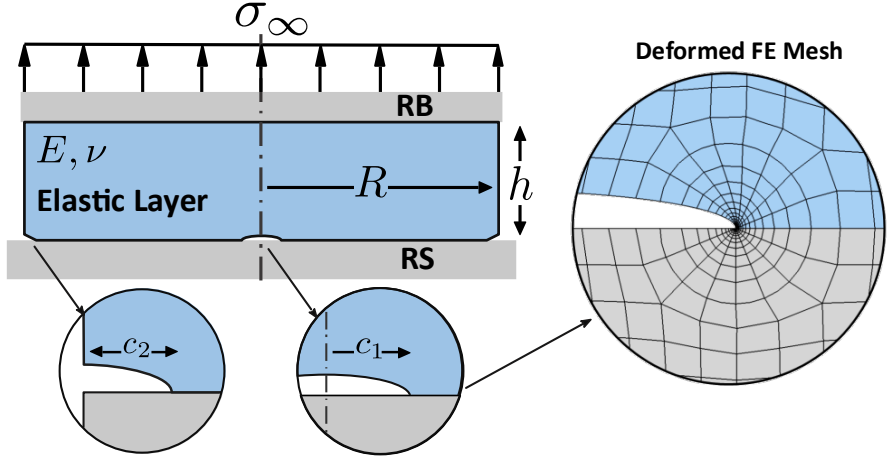

*Figure 1* Sketch of the model system: an elastic disk is perfectly bonded to a rigid backing (RB) and adhered to a rigid substrate (RS). The disk-RS interface has imperfections described by a center penny-shaped crack, $c_1$, and an edge ring-shaped crack $c_2$. The deformed FE Mesh shows how a spider web-shaped mesh around the crack refines the mesh perfectly around the singularity point.

**Results**

We consider three cases: (i) absence of cracks ($c_1 = c_2 = 0$); (ii) presence of only the center crack, ($c_1 \neq 0$, and $c_2 = 0$); (iii) presence of only the edge crack, ($c_1 = 0$, and $c_2 \neq 0$).

The stress field at the interface in the model system sketched in Figure 1 is described as a function of the dimensionless radial coordinate originated from the axis of symmetry $\hat{r} = r/R$, which ranges from 0 (center) to 1 (edge).

*Case (i) Stress distribution in the absence of cracks*

In the absence of cracks, we expect to see a stress singularity at the perimeter of the interface due to no-slip condition. This singularity is comprehensively studied in [31]. As the thickness of the layer decreases, the stress concentration at the edge of the adhesive decreases. However, this also increments the interfacial stress at the center of the disk, as observed by [26]. To confirm this, we performed an asymptotic analysis assuming very small tip thickness ($\hat{h} \ll 1$), perfect incompressibility ($\nu = 0.5$) adapted from [30]. In this case, as detailed in Appendix A, the distribution of normal stress at the interface becomes

$$\sigma_{22} = 2\sigma_\infty(1 - \hat{r}^2) \qquad (9)$$

According to Eq. (9), for a very thin incompressible layer, the normal stress at the center of the interface ($\hat{r} = 0$) is twice the remote stress applied.

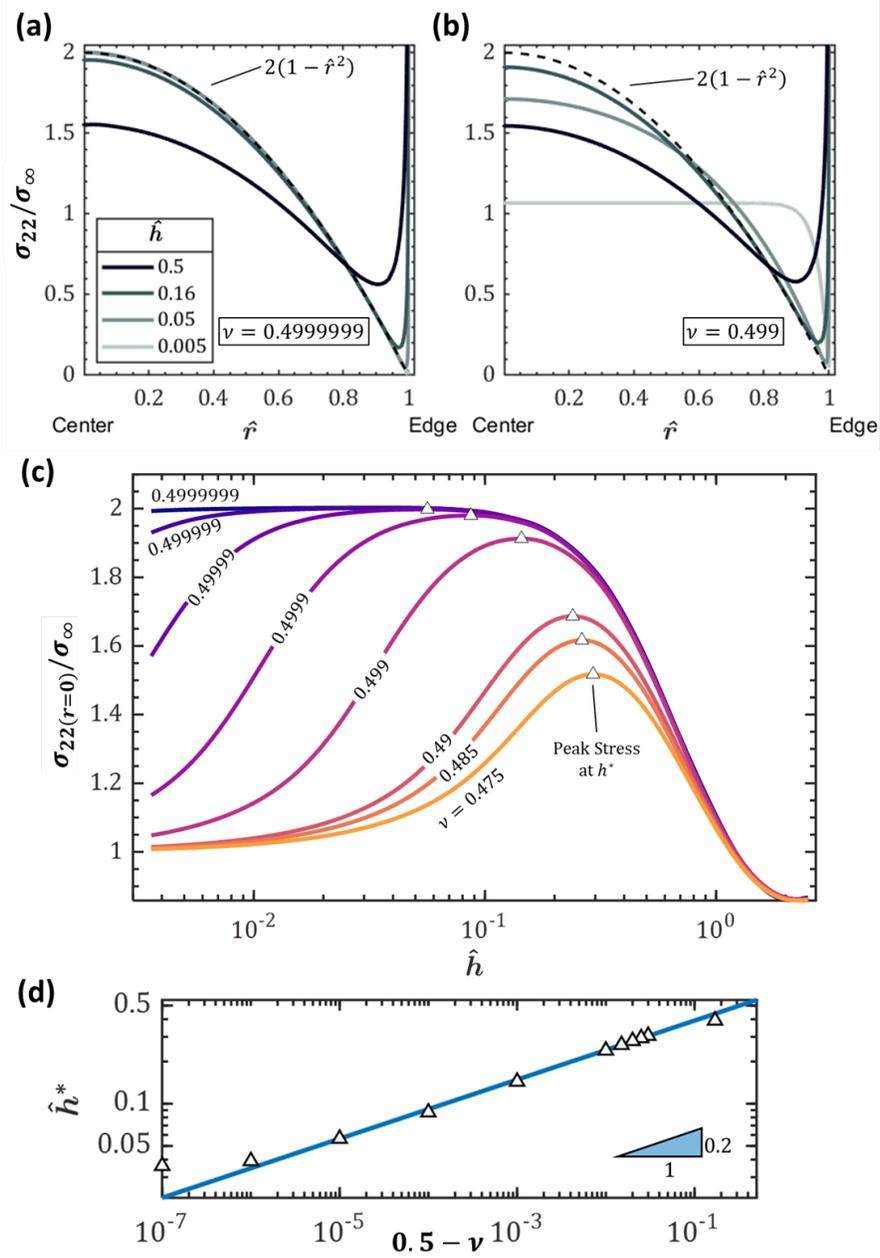

*Figure 2* Distribution of interfacial stress in a circular elastic layer sandwiched between two rigid plates for $\nu = 0.4999999$ (a) and $\nu = 0.499$ (b). The normalized stress at the center of the interface is function of the normalized thickness $\hat{h}$ and Poisson's ratio $\nu$ (c). The stress peak occurs at $\hat{h}^*$, a function of $\nu$ (d).

Figure 2(a) reports the results from our FE analysis for $\nu = 0.4999999$ (the closest condition to incompressibility among the ones we explored). As $\hat{h}$ approaches 0, the interfacial stress distribution approaches that of Eq. (9). For larger thicknesses, the stress concentration at the center of the interface reduces as $\hat{h}$ increases.

Figure 2(b) reports our results for the case of $\nu = 0.499$. A reduction in thickness $\hat{h}$ initially produces an increment of stress concentration at the center of the interface, with the overall stress distribution approaching the solution of Eq. (9). However, for $\hat{h} \leq 0.16$, a reduction in $\hat{h}$ makes the stress distribution more uniform, with the stress concentration at the center disappearing as $\hat{h}$ approaches zero.

Figure 2(c) analyzes the stress concentration at the center of the disk through the ratio $\sigma_{22}(r = 0)/\sigma_\infty$, as a function of $\hat{h}$, with a semilog plot and for various Poisson's ratios. The nearest incompressible case, at $\nu = 0.4999999$, evidences the stress concentration in Eq. (9). For smaller values of $\nu$, the stress concentration presents a peak at $\hat{h} = \hat{h}^*$. For $\hat{h} > \hat{h}^*$, a reduction in $\hat{h}$ produces an increment in stress concentration. For $\hat{h} < \hat{h}^*$, a reduction in $\hat{h}$ produces a reduction in stress concentration with $\sigma_{22}(r = 0)/\sigma_\infty$ approaching unity. The transition thickness of the adhesive, $\hat{h}^*$, is reported in Figure 2(d) as a function of $0.5 - \nu$ with a log-log plot. The linear trend in this plot evidences a power-law correlation between $\hat{h}^*$ and $0.5 - \nu$. Most polymers exhibit $\nu = 0.475 - 0.49$. In this case, as evidenced in Figure 2(c), the stress concentration at the center of the contact region vanishes for $\hat{h} \leq 0.03$.

*Cases (ii), center crack, and (iii), edge crack*

<u>Stress intensity factor</u>

The stress singularities at the tips of the center crack and the edge crack can be described with mixed more fracture mechanics as described in the previous section. In Figures 3 and 4 we report the results for $\nu = 0.4999$. At different Poisson ratios, the results are qualitatively similar.

Figure 3 shows the real and imaginary parts of the dimensionless complex stress intensity factor $\widehat{\boldsymbol{K}}$ described in Eq. (8). The real part, $\widehat{K}_I$, is associated with Mode I fracture, while the imaginary part, $\widehat{K}_{II}$, is associated with Mode II.

Let us first analyze the center crack. For $\hat{h} \geq 0.25$, the crack propagates in pure Mode I, hence $\widehat{K}_{II} \simeq 0$ (the line is not shown in the figure). In this case, a reduction in $\hat{h}$ produces an increment in $\widehat{K}_I$, if $\hat{c} \leq \hat{h}$. For $\hat{h} \geq 1$, $\widehat{K}_I$ is independent of $\hat{c}$ and approaches that of a penny-shaped crack in a semi-infinite space. For $\hat{h} < 0.25$, we have $\widehat{K}_{II} \neq 0$, hence the crack propagates in mixed mode. In this case, a reduction in $\hat{h}$ produces a reduction in $\widehat{K}_I$, suggesting a confinement-induced strengthening.

The edge crack propagates in mixed mode for any $\hat{h}$. $\widehat{K}_I$ is larger than that for center crack and the values for $\widehat{K}_I$ and $\widehat{K}_{II}$ match the results obtained by [31]. For $\hat{c} \leq 0.25$, a reduction in $\hat{h}$ produces a reduction of both $\widehat{K}_I$ and $\widehat{K}_{II}$.

Considering the phase angle $\psi$ plot in Figure 3, we can observe here that a smaller tip thickness provides more mode mixing for both cracks. For center crack, $\psi$ approaches zero as $\hat{c}$ reduces below ~ $\hat{h}/2$, hence providing pure Mode I propagation. For $\hat{c} > \hat{h}/2$, $\psi$ grows proportional to $\hat{c}$, evidencing mode mixing for cracks as large as the tip thickness.

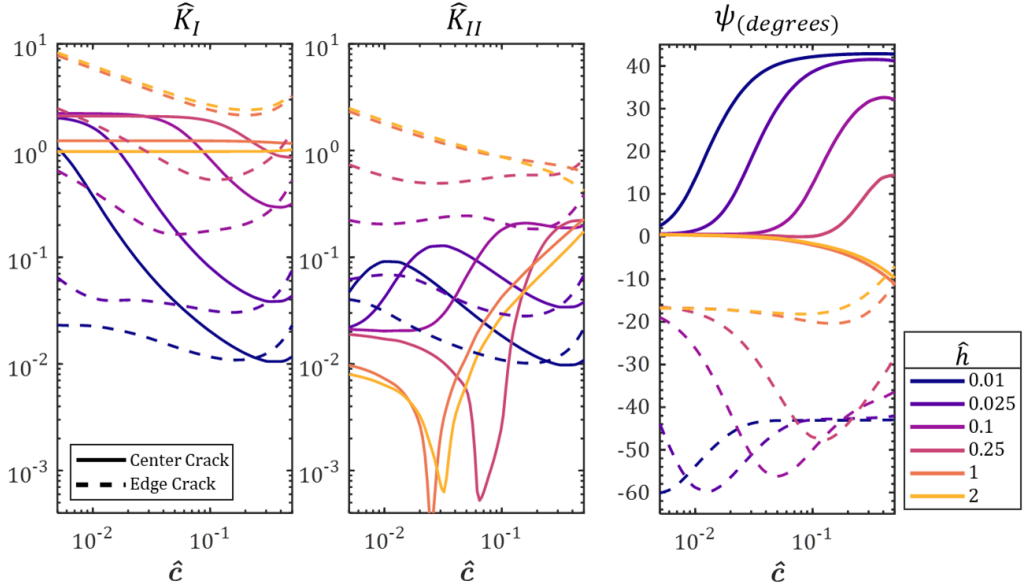

*Figure 3* Real ($\widehat{K}_I$) and imaginary ($\widehat{K}_{II}$) parts and phase angle ($\psi$) of the dimensionless stress intensity factor for center crack (solid lines) and edge crack (dashed lines) at various crack size $\hat{c}$, tip thickness $\hat{h}$, and $\nu = 0.4999$.

For edge crack, mode mixing is always present due to no slip condition. $\psi$ approaches $-20°$ as $\hat{c}$ reduces below ~ $\hat{h}/20$. For $\hat{c} < \hat{h}/2$, $\psi$ grows in magnitude with $\hat{c}$ up to a peak at $\hat{c} \approx \hat{h}/2$. Further increment of $\hat{c}$ produces a reduction in magnitude in $\psi$ toward an asymptotic value that is inversely proportional to $\hat{h}$.

Energy Release Rate

The strain energy release rate $G$ is the driving force for crack propagation, while the energetic cost for crack propagation is the work of adhesion $W_A$. Detachment occurs when

$G \geq W_A$ (10)

since crack propagation is spontaneous. $W_A$ is proportional to the strength of the interfacial bonds, and we assume it is a constant of equal value for both center and edge cracks. $G$ must be calculated separately for center and edge cracks, with the largest one giving the preferred detachment mechanism, from Eq. (10).

A small center crack ($\hat{c} \ll 1$) can be treated as a penny-shaped crack in an infinite bi-material interface. For this case, the energy release rate computes to [39]

$$G = \frac{\left|2\sigma_\infty\sqrt{c}\frac{\Gamma(1+i\epsilon)}{\Gamma(\frac{1}{2}+i\epsilon)}(2\hat{c})^{i\epsilon}\right|^2}{2E^*\cosh(\pi\epsilon)} \tag{11}$$

with $\Gamma$ the gamma function $\Gamma(z) = \int_0^\infty t^{z-1} e^{-t} dt$, where $z$ is a complex number with positive real part. In our case, we have $\epsilon \approx 0$, hence Eq. (11) simplifies. For thin adhesive thicknesses, the stress transmitted across the interface in the absence of cracks is non-uniform. Hence the remote stress $\sigma_\infty$ in Eq. (11) should be replaced with the stress transmitted at the center of the uncracked interface, $\sigma_{zz}(r=0)$, extract from Figure 2(c); thus,

$$G = \frac{2\sigma_{zz}^2(r=0)}{\pi E^*}c \tag{12a}$$

giving

$$\hat{G} = \frac{2}{\pi}\left(\frac{\sigma_{zz}(r=0)}{\sigma_\infty}\right)^2 \hat{c} \tag{12b}$$

for the dimensionless energy release rate

$$\hat{G} = \frac{GE'}{\sigma_\infty^2 R} \tag{13}$$

Based on Figure 2(c), the ratio $\sigma_{zz}(r=0)/\sigma_\infty$ varies between 1 and 2, hence $2/\pi < \hat{G}/\hat{c} < 8/\pi$.

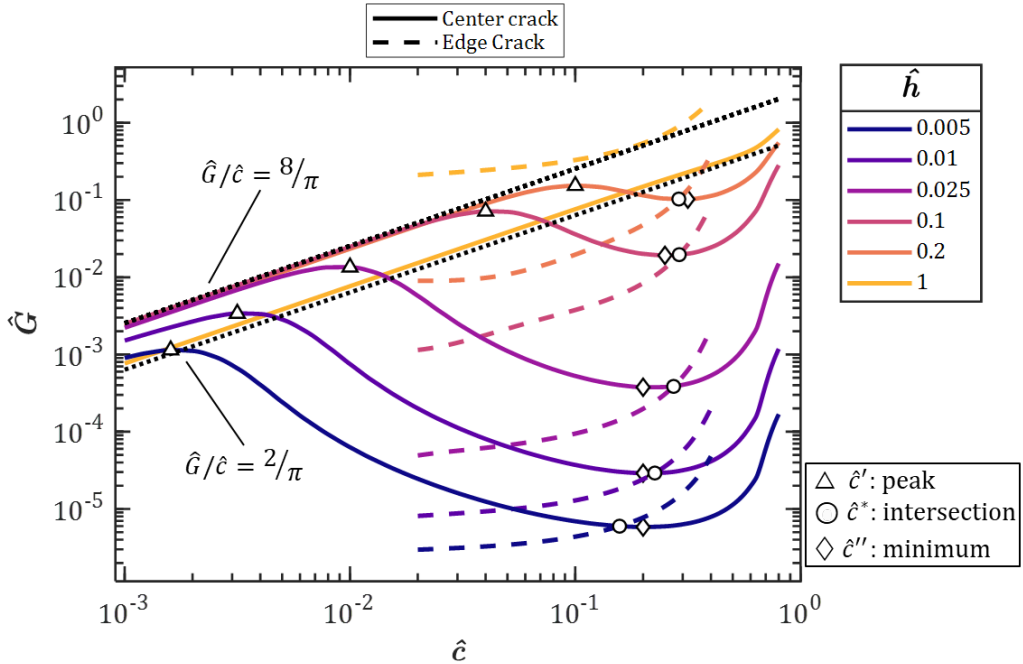

*Figure 4* Dimensionless energy release rate $\hat{G}$ versus dimensionless crack size $\hat{c}$ for center crack (solid lines) and edge crack (dashed lines) for various $\hat{h}$ and $\nu = 0.4999$. For center crack, the points of local maxima are marked with triangles, while the points of local minima are marked with diamonds. The circles also correspond to the points of intersection between center crack and edge crack.

Figure 4 shows the normalized energy release rate $\hat{G}$, as a function of $\hat{c}$ for center crack (solid line) and edge crack (dashed line), for various $\hat{h}$ and $\nu = 0.4999$. For center crack and very small $\hat{c}$, $\hat{G}$ grows proportionally to $\hat{c}$ until a local maximum at $\hat{c} = \hat{c}'$ (triangles). For $\hat{c} > \hat{c}'$, further crack growth produces a reduction in $\hat{G}$ until the global minimum at $\hat{c}''$ (diamonds). At some point the solid line (center crack) intersects the dashed line (edge crack) at $\hat{c} = \hat{c}^*$ (circles), thereby defining the transition between center crack domination and edge crack domination. $\hat{c} < \hat{c}^*$ favors center crack propagation, while $\hat{c} > \hat{c}^*$ favors edge crack propagation. As observed in the figure, the intersection points correspond approximatively with the points of minimum for center crack, hence $\hat{c}^* \approx \hat{c}''$. It is worth notice that in the curve associated with $\hat{h} = 1$, there is no intersection point and edge crack propagation is always favored. This is generally true for $\hat{h} \geq 0.8$ and explains why long compliant cylindrical pillars always detach from the perimeter.

Figure 4 can be used to study the stability of center crack propagation. Unstable propagation relies on Eq. (10) and, since crack propagation increments $c$, this equation can be differentiated to give $\partial G/\partial c \geq \partial W_A/\partial c$. Because the work of adhesion is commonly uniformly distributed across the interface, $\partial W_A/\partial c = 0$ and $\partial G/\partial c \geq 0$ provides the condition for continuous propagation. Conversely, $\partial G/\partial c < 0$ prevents further propagation as $G$ reduces below $W_A$, unless a larger load is applied. This provides the condition for crack stability, which rewrites in dimensionless form as

$$\frac{\partial \hat{G}}{\partial \hat{c}} < 0 \tag{14}$$

Eq. (14) is satisfied for $\hat{c}' < \hat{c} < \hat{c}''$, the stable region between the local maximum and minimum of $\hat{G}$. Small cracks satisfying $\hat{c} \leq \hat{c}'$ propagate unstably until reaching the stable region and then stop. Conversely, cracks as large as $\hat{c} \geq \hat{c}''$ propagate unstably until complete detachment. Because $\hat{c}'' \approx \hat{c}^*$, the condition $\hat{c} \leq \hat{c}^*$ ensures center crack domination and stable propagation.

Figure B1, in *Appendix B*, reports the same results as Figure 4 but for different Poisson's ratios.

In Figure 5(a) we correlate $\hat{c}'$ (triangles) and $\hat{c}^*$ (circles) with $\hat{h}$ in a log-log plot, for various $\nu$. For clarity, we omit here the correlation between $\hat{c}''$ and $\hat{h}$, however, this can be estimated from the relation $\hat{c}'' \approx \hat{c}^*$, and accurately extrapolated from Figure 4 and Figure B1, in *Appendix B*. Figure 8 shows a linear correlation between $\hat{c}'$ and $\hat{h}$, with small dependence from $\nu$. $\hat{c}^*$ evidences a square root correlation with $\hat{h}$ before reaching a plateau at $\hat{c}^*_{max} \simeq 0.28$. The transition between the square root correlation and the plateau is controlled by $\nu$, with the extreme case of $\nu = 0.499999$ exhibiting only constant $\hat{c}^* = \hat{c}^*_{max}$. $\hat{c} \geq \hat{c}^*_{max}$ provides then unstable edge crack propagation for any $\hat{h}$ and any $\nu$.

The combination of Eq. (10) and (13) provides the condition $\hat{G} \geq W_A E^*/\sigma_\infty^2 R$ for crack propagation. This rewrites to $\sigma_\infty \geq \sigma_c = \sqrt{W_A E^*/\hat{G} R}$, with $\sigma_c$ the detachment strength of the adhesive interface. The latter can be written in dimensionless form as

$$\hat{\sigma}_c = \frac{\sigma_c \sqrt{R}}{\sqrt{W_A E^*}} = \frac{1}{\sqrt{\hat{G}}} \tag{15}$$

From Eq. (15) we can intuitively observe that the lowest $\hat{G}$, $\hat{G}'' = \hat{G}(\hat{c}'')$, provides the highest adhesive strength, i.e. the theoretical limit in adhesive strength of the interface

$$\hat{\sigma}_c^* = \frac{1}{\sqrt{\hat{G}''}} \tag{16}$$

Figure 5(b) shows the correlation between $\hat{\sigma}_c^*$ and $\hat{h}$ at various $\nu$. In this figure, we can observe an inverse proportionality between $\hat{\sigma}_c^*$ and $\hat{h}$, for any $\nu$. For $\nu \geq 0.499999$, the curve matches the power-law correlation $\hat{\sigma}_c^* \simeq 0.25\, \hat{h}^{-1.44}$ which, by inverting Eq. (15), gives

$$\sigma_c^* \simeq \frac{0.25}{\sqrt{h}} \left(\frac{R}{h}\right)^{0.94} \sqrt{W_A E^*} \tag{17}$$

the closed-form estimation of the maximum strength for an incompressible adhesive.

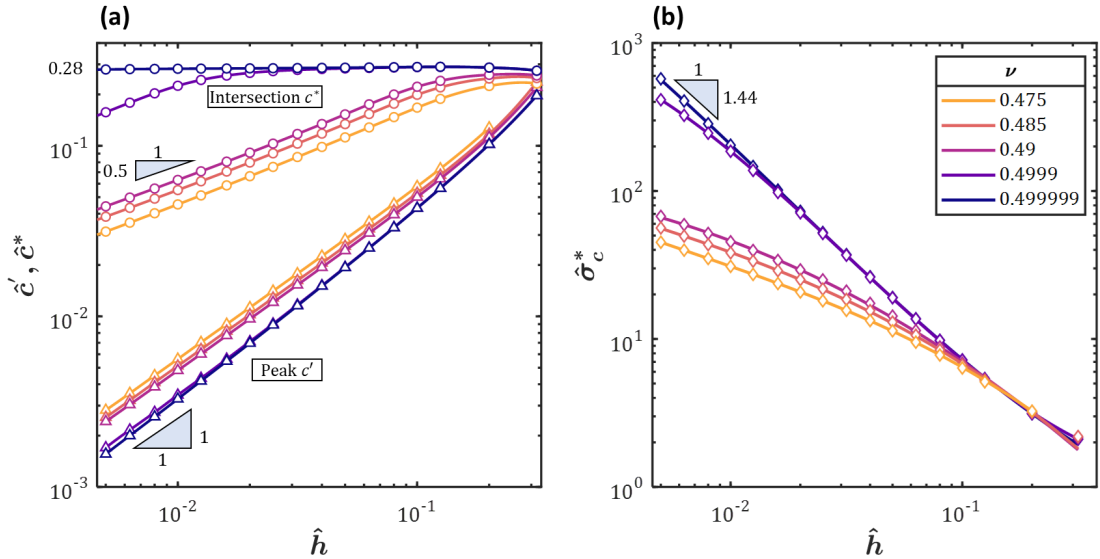

*Figure 5(a)* Crack sizes $\hat{c}'$ (triangles) and $\hat{c}^*$ (circles), and *(b)* Maximum dimensionless detachment strength $\hat{\sigma}_c^*$ versus $\hat{h}$ for various $\nu$.

**Discussion and Conclusions**

The detachment strength of a axisymmetric bi-material adhesive can be predicted from the results shown in Figures 3-5 and B1 (*Appendix B*). For a given dimensionless tip thickness, $\hat{h}$, dimensionless statistical interfacial defect, $\hat{c}$, and Poisson's ratio $\nu$, Figures 4 and B1 provide the dimensionless energy release rate $\hat{G}$ for both center and edge cracks, with the largest one controlling detachment. From this, one can then derive the interfacial strength $\sigma_c = \sqrt{W_A E^*/\hat{G} R}$, using Eq. (15), with $\hat{G}$ the maximum between center crack and edge crack. This prediction is accurate if $\hat{c} \geq \hat{c}^*$, where the edge crack dominates, and condition $\sigma_\infty \geq \sigma_c$ provides unstable crack propagation. If $\hat{c} < \hat{c}^*$, however, center crack dominates and, in particular when $\hat{c} < \hat{c}''$, crack propagation is stable, hence the previous prediction is too conservative. In this case, unstable crack propagation will not occur until $\sigma_\infty \geq \sigma_c^* = \hat{\sigma}_c^* \sqrt{W_A E^*/R}$, with $\hat{\sigma}_c^*$ extracted from Figure 5(b), as a function of $\hat{h}$ and $\nu$. $\sigma_c^*$ ultimately constitutes the maximum theoretical strength of the adhesive. For a Poisson's ratio $\nu \geq 0.499999$, $\sigma_c^*$ is provided by Eq. (17) in closed form, suggesting that adhesives with smaller tip thickness $h$ and larger radius $R$ can achieve higher strength. The same principle applies qualitatively also for smaller Poisson's ratios than 0.499999. It is worth note that, while a smaller $h$ can in principle increment $\sigma_c^*$, it might reduce the ability of the adhesive to conform to the roughness of the RS, thereby reducing the effective work of adhesion $W_A$. In our analysis, we neglect surface roughness and assume perfect conformation between the adhesive and the adhered RS.

The maximum theoretical strength $\sigma_c^*$ is only available under center crack domination and stable propagation, i.e. $\hat{c} \leq \hat{c}^*$ and $\hat{c} \leq \hat{c}''$. This leads to the condition $R \geq c/\hat{c}^*$ and $R \geq c/\hat{c}''$, providing a scaling law between the radius of the adhesive and the size of the center crack. The latter can be estimated from the maximum size of an interfacial flaw, created by manufacturing defects or other imperfections. For the nearly incompressible case of $\nu = 0.499999$, as observed in Figure B1, $\hat{c}^* \leq \hat{c}''$, hence $R \geq c/\hat{c}^*$ is sufficient to ensure the aforementioned conditions. As observed in Figure 5(a), $\hat{c}^* \leq 0.28$ for any $\hat{h}$, hence the condition $R \geq 3.54\ c$ ensures maximum theoretical strength for incompressible adhesives. For a smaller Poisson ratio, one should obtain $\hat{c}^*$ and $\hat{c}''$ from Figure 4 and B1 and compare them against $R$ and $c$.

Our analysis considers crack-like interfacial defects statistically located everywhere with equal probability. The interfacial stress distribution, in the absence of cracks, shows the largest stress concentrations at the center and the edge of the interface. This provided the evidence of center crack and edge crack propagation as the predominant detachment mechanisms, hence affecting the adhesive strength. However, we did not consider the hypothesis of multiple cracks simultaneously present at the interface. If we assume all cracks are small, i.e. $\hat{c} \ll 1$, and their maximum spacing is much larger than their maximum size, we can consider their propagation to be independent. Under these conditions, the strength of the adhesive is still statistically controlled by the two predominant mechanisms, center and edge crack propagation. If two or more cracks become too large or too close to each other, so that the hypothesis of crack-independence is no longer satisfied, one needs to account for their mutual interaction. The adhesive strength in this case is lower than the theoretical limit provided because crack interaction promotes unstable propagation. This conclusion can be extrapolated by Figure 4, where larger center cracks become unstable as their tip gets closer to the edge of the interface.

We also assume that a small crack would not propagate unstably from any locations of the interface during stable propagation of a center crack. The asymptotic analysis of interfacial stress in *Appendix A* shows that for the case of a center crack propagation within an incompressible adhesive, the interfacial normal stress transmitted across the interface reaches a local maximum at a distance $\hat{r}^*$ from the center satisfying $\hat{c} \leq \hat{r}^* \leq 1$. Figure A4 shows that for $\hat{c} \geq 0.5$, the stress concentration at $\hat{r}^*$ is larger than that initially at the center of the interface (see Figure 2). From this observation, one might consider that unstable crack propagation could occur at $\hat{r}^*$ before center crack propagation becomes unstable. However, Figures 4-5 shows that center crack propagation is unstable for $\hat{c} \geq 0.28$, supporting our hypothesis of center crack domination. This is also in line with findings of [20] comparing the stability of edge and center crack propagation experimentally.

The theoretical maximum strength we estimated is purely based on linear elastic fracture mechanics (LEFM) and provides the interfacial toughening created by the confinement of the soft adhesive tip, sandwiched between the RB and the RS. It is worth noticing that the interfacial adhesive strength is also limited by the cohesive strength created by van der Waals bonds at the interface, $\sigma_{c,\text{vdW}}$, which is independent of the geometry of the adhesive. Hence, the maximum adhesive strength is the lowest value between $\sigma_c^*$ and $\sigma_{c,\text{vdW}}$. Under the condition $c \leq \hat{c}^* R$, $\sigma_c^*$ is independent of $c$, hence the adhesive is flaw tolerant.

The ambient pressure can also be a player, in the adhesive strength during center crack propagation by creating a suction effect inside the penny-shaped crack regarding the pressure drop during its propagation. This suction increases the adhesive strength yet its effectiveness as experimented by [33] also depends on the air leakage into the cavities due to interfacial defects.

Our LEFM-based analysis is quasi-static. It relies on the assumption of small displacements and small strains and no rate-dependent material behavior. Due to the small magnitude of van der Waals forces, most adhesives do not experience significant strain at the detachment, thus justifying our assumptions. Some adhesives, however, might experience significant strain during detachment. In this case, a nonlinear elastic analysis is required and our prediction only provides a first-order approximation. Rate-dependent dissipative mechanisms such as viscous dissipation can reduce the strain energy release rate during crack propagation, thereby toughening the interface. This contribution can be accounted for by describing a rate-dependent work of adhesion $W_a$, which accounts for the toughening contribution. However, in this case a rate dependent elastic response can also create a stress redistribution across the interface, thereby requiring a more detailed analysis.

Our method is based on the stress analysis of the uncracked interface for various values of the geometrical characteristics of the adhesive (in this case the soft layer thickness) to identify regions affected by stress concentration. Then it investigates the evolution of the strain energy release rate due to the propagation of cracks located in the stress concentration regions. Finally, it identifies the main detachment mechanisms and their crack propagation stability to compute the maximum adhesive strength. This approach can be extended to other adhesive geometries having a relatively small number of geometric characteristics.

**Acknowledgments**
We thank Attila Kossa (Budapest University of Technology and Economics) for helpful conversations. This work was supported by the Natural Sciences and Engineering Research Council of Canada (NSERC) (RGPIN-2017-04464).

# Appendix A - Asymptotic analysis of interfacial stress distribution in the presence of a center crack.

The interfacial stress distribution of a cylindrical bi-material adhesive having a thin compliant tip has been investigated [30]. Here we extend that solution considering the presence of a penny-shaped crack at the center of the adhesive interface. First, we reproduce the solution for the case of an uncracked interface.

Consider a compliant incompressible disk of radius $R$, thickness $h$, and shear modulus $\mu$ adhered to a rigid substrate at its bottom and perfectly bonded to a rigid stalk at its top. The condition of a thin tip implies $h \ll R$. A remote stress $\sigma_\infty$ is applied to separate the interface, creating a displacement $\Delta$. We consider no slip condition between the adhesive and the rigid substrate, and between the adhesive and the rigid stalk. The lateral surface of the disk is traction free. Adopting polar coordinates $r, \theta, z$, we describe the displacement field in the disk with $u_r, u_\theta, u_z$, assuming $u_\theta = 0$ due to axisymmetry. Figure A1 provides a sketch of our model. We consider the cylindrical portion of material above the center crack to be unloaded, under the hypothesis of $c > h$.

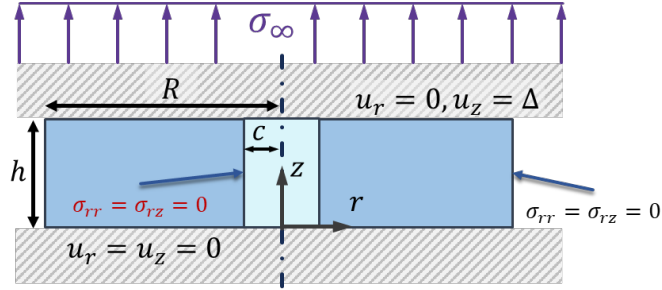

*Figure A1-* Sketch of the thin tip bi-material adhesive having a penny-shaped crack at its center. The cyan cylinder at the center indicates the unloaded material above the crack.

Let us now introduce the dimensionless disk thickness as $\hat{h} = h/R$, with our hypothesis of thin tip imposing $\hat{h} \ll 1$. The dimensionless polar coordinates are then $\hat{r} = r/R$ and $\hat{z} = z/h$, with the dimensionless displacements being $\hat{u}_r = u_r/\Delta$ and $\hat{u}_z = u_z/\Delta$. By adopting the nominal stress $\Sigma = \mu R^2 \Delta / h^3$, we finally obtain the dimensionless stresses $\hat{\sigma}_{ij} = \sigma_{ij}/\Sigma$. The nominal stress is comparable in magnitude with the remote stress $\sigma_\infty$, as will be demonstrated at the end of this section.

Expanding the stress components in Taylor's series with respect to the variable $\hat{h}$ we have

$$\hat{p} = \hat{p}^{(0)} + \hat{p}^{(1)}\hat{h} + O(\hat{h}^2)$$

$$\hat{\sigma}_{rr} = -\hat{p}^{(0)} - \hat{p}^{(1)}\hat{h} + O(\hat{h}^2)$$
$$\hat{\sigma}_{\theta\theta} = -\hat{p}^{(0)} - \hat{p}^{(1)}\hat{h} + O(\hat{h}^2)$$
$$\hat{\sigma}_{zz} = -\hat{p}^{(0)} - \hat{p}^{(1)}\hat{h} + O(\hat{h}^2)$$
$$\hat{\sigma}_{rz} = \hat{\sigma}_{rz}^{(1)}\hat{h} + O(\hat{h}^2)$$

(A1)

In this equation, $\hat{p} = p/\Sigma$ is the dimensionless hydrostatic pressure, with $p = -(\sigma_{rr} + \sigma_{\theta\theta} + \sigma_{zz})/3$. According to Eq. (A1), the deviatoric components of the stress tensor are higher-order with respect to the hydrostatic component. Expanding now also the displacements, in the same fashion as in Eq. (A1), we have

$$\hat{u}_r = \hat{u}_r^{(-1)} \frac{1}{\hat{h}} + \hat{u}_r^{(o)} + O(\hat{h})$$
$$\hat{u}_z = \hat{u}_r^{(o)} + \hat{u}_r^{(1)}\hat{h} + O(\hat{h}^2) \tag{A2}$$

This equation evidences the predominance of the radial displacement over the axial one [30]. This is due to incompressibility. Hooke's law, equilibrium and compatibility are satisfied for Eq. (A1) and (A2) if the leading terms of $\hat{p}$ become

$$\hat{p}^{(0)} = 3\,\hat{r}^2 + A\ln(\hat{r}) + B$$
$$\hat{p}^{(1)} = 0 \tag{A3}$$

with $A$ and $B$ integration constants. Substituting Eq. (A3) into (A1) and imposing traction free boundary conditions $\hat{\sigma}_{rr} = 0$ for $\hat{r} = 1$, and finite stress at the center, we have $A = 0$ and $B = -3$.

To equilibrate the interfacial stress with the remotely applied $\sigma_\infty$, one has to impose

$$\sigma_\infty(\pi R^2) = 2\pi R^2 \Sigma \int_0^1 \hat{\sigma}_{zz}\hat{r}d\hat{r} \tag{A4}$$

Substituting Eq. (A3) into (A1) and then into (A4), we finally have $\Sigma = 2\sigma_\infty/3$, and hence

$$\sigma_{zz} = 2\sigma_\infty(1 - \hat{r}^2) + O(\hat{h}^2) \tag{A5}$$

Eq. (A5) shows a parabolic distribution of normal interfacial stress with a global maximum of $2\sigma_\infty$ at the center of the interface ($r = 0$). This is in agreement with our FE simulations in Figure 2(a) for $\hat{h} = 0.005$.

Consider now the presence of a circular (penny-shaped) crack of dimensionless size $\hat{c} = c/R$ at the center of the interface. For $0 \leq \hat{r} < \hat{c}$, $\sigma_{zz} = 0$ at the interface. This traction-free condition can, however, extend through all the thickness of the layer for the case of $\hat{h} < \hat{c}$, under the hypothesis of a thin layer ($\hat{h} \ll 1$). Figure A1 shows the traction-free cylindrical domain above the center crack, while Figure A2 shows the contour plot of the stress distribution obtained from finite element analysis (FEA).

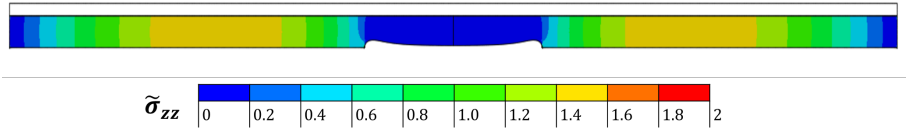

*Figure A2*- Stress distribution of the thin elastic adhesive layer in the presence of a center crack form finite element analysis (FEA). The dark blue contour indicates traction free regions ($\tilde{\sigma}_{zz} = \sigma_{zz}/\sigma_\infty = 0$).

Assume now the material inside the traction-free domain to have zero stress in all components, hence $\hat{\sigma}_{zz} = \hat{\sigma}_{rr} = \hat{\sigma}_{rz} = 0$ for $\hat{r} \leq \hat{c}$ (all the other terms are zero due to axisymmetry). The condition $\hat{\sigma}_{rr} = 0$ for $\hat{r} = \hat{c}$ gives $A = 3(1 - \hat{c}^2)/\ln \hat{c}$ and $B = -3$, which substituted into Eq. (A3) give

$$\hat{\sigma}_{zz} = 3\left[1 - \hat{r}^2 - (1 - \hat{c}^2)\frac{\ln \hat{r}}{\ln \hat{c}}\right] + O(\hat{h}^2) \tag{A6}$$

Substituting Eq. (A6) into (A4) we have $3[(1 - \hat{c}^2)^2 + (1 - \hat{c}^4)\ln \hat{c}]\Sigma = 2\sigma_\infty \ln \hat{c}$, which finally gives

$$\sigma_{zz} = 2\sigma_\infty \frac{(1-\hat{r}^2)\ln \hat{c} + (1-\hat{c}^2)\ln \hat{r}}{(1-\hat{c}^4)\ln \hat{c} + (1-\hat{c}^2)^2} + O(\hat{h}^2) \tag{A7}$$

For the case of $\hat{c} = 0$, Eq. (A7) gives (A5). The interfacial stress distribution described by Eq. (A7) neglects the stress singularity at the crack tip, hence only gives a macroscopic description of the interfacial stress for the asymptotic case of $\hat{h} \ll 1$. *Figure A3* shows a comparison between the stress distribution given by Eq. (A5) and (A7) (solid lines) and that given by FE analysis ( lines). The latter is to be considered more accurate since it accounts for the stress singularity at the crack tip. The stress here is normalized as $\tilde{\sigma}_{zz} = \sigma_{zz}/\sigma_\infty$. Our FE simulations assumed $\hat{h} = 0.01$, $\nu = 0.4999999$, and $c$ varying between 0.002 and 0.2. As shown in the figure, for $\hat{c} < \hat{h}$ the FE results are in good agreement with Eq. (A5), while for $\hat{c} \geq \hat{h}$ the FE results are in agreement with Eq. (A7).

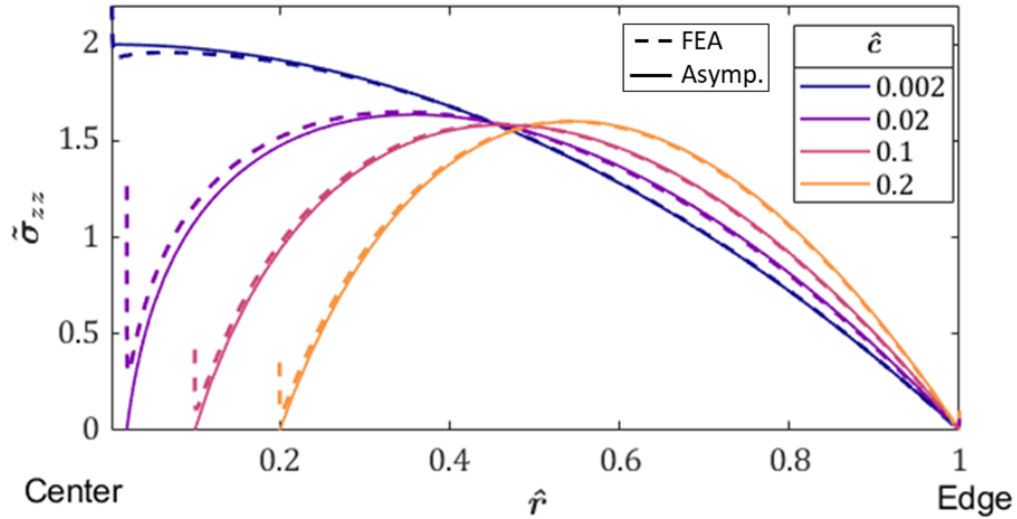

*Figure A3*- Interfacial stress distribution provided by the asymptotic solution in Eq. (A5), for $\hat{c} = 0.002$, and (A7), for $\hat{c} = 0.02 - 0.2$ (solid) in comparison to interfacial stress distribution obtained from FEA (dashed) for a layer having $\hat{h} = 0.01$, and $\nu = 0.4999999$.

From *Figure A3* we can observe a stress peak at $\hat{r} = \hat{r}^*$, with $\hat{c} \leq \hat{r}^* \leq 1$, due to the presence of the center crack, suggesting the possibility of crack propagation at an intermediate location between the center and the perimeter of the contact region. The position $\hat{r}^*$ and value of the stress concentration $\tilde{\sigma}_{zz}^* = \sigma_{zz}^*/\sigma_\infty$ evidenced in Figure A3 can be calculated from the stationarity of Eq. (A7). This gives

$$\hat{r}^* = \sqrt{\frac{\hat{c}^2-1}{2\ln\hat{c}}} \tag{A8}$$

and

$$\tilde{\sigma}_{zz}{}^* = \frac{1+2\ln\hat{c}-\hat{c}^2+(1-\hat{c}^2)[\ln(\hat{c}^2-1)-\ln(2\ln\hat{c})]}{(1-\hat{c}^4)\ln\hat{c}+(1-\hat{c}^2)^2} \tag{A9}$$

Figure A4 reports the plot of $\hat{r}^*$ and $\tilde{\sigma}_{zz}{}^*$ versus $\hat{c}^*$ derived from Eq. (A8) and (A9), respectively. For $\hat{c} \leq 0.5$, $\tilde{\sigma}_{zz}{}^* \leq 2$ and $\hat{r}^* \leq 0.7$, hence the stress concentration is smaller that at the center of the uncracked interface, from Figure A3. If $\hat{c} > 0.5$, we have $\tilde{\sigma}_{zz}{}^* > 2$ and $\hat{r}^* > 0.7$, which evidence a larger stress concentration at $\hat{r}^*$ than at the center. However, crack propagation at $\hat{r} = \hat{r}^*$ is unlikely to be favored since the center crack is already significantly extended and a similar size crack at $\hat{r}^*$ would render the entire interface completely debonded.

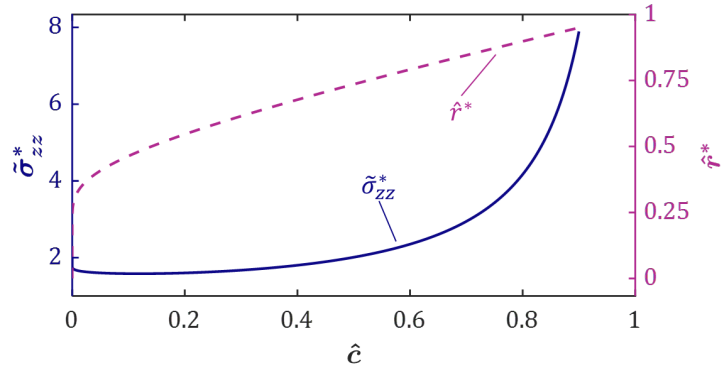

*Figure A4-* Stress concentration $\tilde{\sigma}_{zz}{}^*$ at $\hat{r}^*$ as functions of center crack size $\hat{c}$.

# Appendix B – Influence of Poisson's ratio on the energy release rate of center and edge interfacial cracks

Figure 4 provides the dimensionless energy release rate $\hat{G} = GE'/\sigma_\infty^2 R$ versus $\hat{c} = c/R$ for various dimensionless layer thicknesses $\hat{h} = h/R$ and Poisson's ratio $\nu = 0.4999$. Here we report in Figure B1 the same plots for various $\nu$. In this figure we can observe a larger influence of Poisson's ratio for smaller $\hat{h}$.

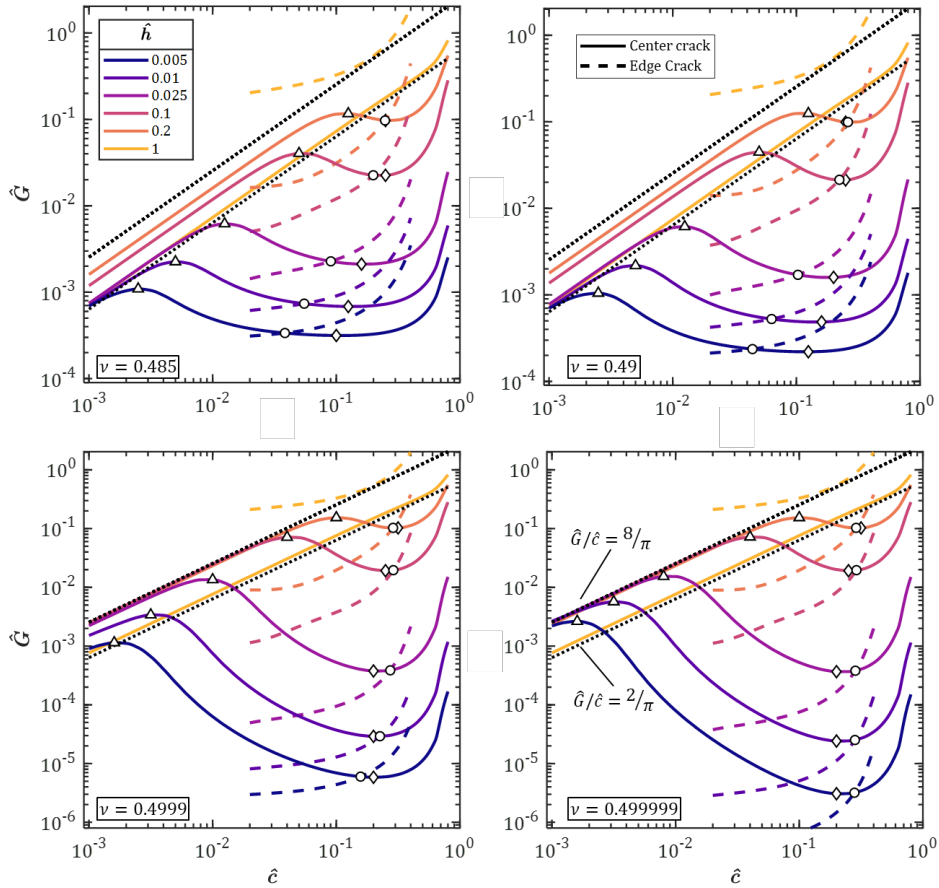

*Figure B1*- Dimensionless energy release rate $\hat{G}$ versus dimensionless crack size $\hat{c}$ for various dimensionless tip thicknesses $\hat{h}$ and Poisson's ratios $\nu$.